\documentclass{emulateapj}

\newcommand{\Kepler}{{\it Kepler}}

\newcommand{\ron}{\color{black}}
  
\newcommand{\thisstar}{WASP-47}
\newcommand{\thisfirstplanet}{WASP-47\,b}
\newcommand{\thissecondplanet}{WASP-47\,e}
\newcommand{\thisthirdplanet}{WASP-47\,d}
\newcommand{\thisfourthplanet}{WASP-47\,c}

\newcommand{\mearth}{M$_\oplus$}
\newcommand{\rearth}{R$_\oplus$}
\newcommand{\msun}{M$_\odot$}
\newcommand{\rsun}{R$_\odot$}
\newcommand{\feh}{[Fe/H]}

\newcommand{\meh}{[M/H]}

\newcommand{\ldone}{0.396}
\newcommand{\uldone}{0.018}
\newcommand{\ldtwo}{0.423}
\newcommand{\uldtwo}{0.017}
\newcommand{\rhost}{0.999}
\newcommand{\urhost}{0.014}
\newcommand{\rprstb}{0.10193}
\newcommand{\urprstb}{0.00021}
\newcommand{\arstb}{9.702}
\newcommand{\uarstb}{0.044}
\newcommand{\inclb}{88.98}
\newcommand{\uinclb}{0.20}
\newcommand{\impb}{0.173}
\newcommand{\uimpb}{0.032}
\newcommand{\rplb}{12.63}
\newcommand{\urplb}{0.15}
\newcommand{\perplb}{4.1591289}
\newcommand{\uperplb}{0.0000042}
\newcommand{\ttransitb}{2457007.932132}
\newcommand{\uttransitb}{0.000021}
\newcommand{\rprste}{0.01461}
\newcommand{\urprste}{0.00013}
\newcommand{\arste}{3.205}
\newcommand{\uarste}{0.014}
\newcommand{\incle}{85.98}
\newcommand{\uincle}{0.75}
\newcommand{\impe}{0.224}
\newcommand{\uimpe}{0.041}
\newcommand{\rple}{1.810}
\newcommand{\urple}{0.027}
\newcommand{\perple}{0.789592}
\newcommand{\uperple}{0.000012}
\newcommand{\ttransite}{2457011.34861}
\newcommand{\uttransite}{0.00033}
\newcommand{\rprstd}{0.02886}
\newcommand{\urprstd}{0.00016}
\newcommand{\arstd}{16.268}
\newcommand{\uarstd}{0.074}
\newcommand{\incld}{89.32}
\newcommand{\uincld}{0.23}
\newcommand{\impd}{0.192}
\newcommand{\uimpd}{0.065}
\newcommand{\rpld}{3.576}
\newcommand{\urpld}{0.046}
\newcommand{\perpld}{9.03077}
\newcommand{\uperpld}{0.00017}
\newcommand{\ttransitd}{2457006.36931}
\newcommand{\uttransitd}{0.00039}

\newcommand{\mplb}{363.1}
\newcommand{\umplb}{7.3}
\newcommand{\rhob}{0.993}
\newcommand{\urhob}{0.021}
\newcommand{\gb}{22.33}
\newcommand{\ugb}{0.27}
\newcommand{\mple}{6.83}
\newcommand{\umple}{0.66}
\newcommand{\rhoe}{6.35}
\newcommand{\urhoe}{0.64}
\newcommand{\gpe}{20.5}
\newcommand{\ugpe}{2.0}
\newcommand{\mpld}{13.1}
\newcommand{\umpld}{1.5}
\newcommand{\rhod}{1.58}
\newcommand{\urhod}{0.18}
\newcommand{\gd}{10.1}
\newcommand{\ugd}{1.1}
\newcommand{\mplc}{398.2}
\newcommand{\umplc}{9.3}
\newcommand{\eccc}{0.296}
\newcommand{\ueccc}{0.017}
\newcommand{\omegac}{112.4}
\newcommand{\uomegac}{4.8}
\newcommand{\perplc}{588.5}
\newcommand{\uperplc}{2.4}
\newcommand{\ttransitc}{2457763.4}
\newcommand{\uttransitc}{4.9}
\newcommand{\kb}{140.64}
\newcommand{\ukb}{0.44}

\newcommand{\kd}{3.93}
\newcommand{\ukd}{0.43}
\newcommand{\ke}{4.61}
\newcommand{\uke}{0.44}
\newcommand{\mst}{1.040}
\newcommand{\umst}{0.031}
\newcommand{\rst}{1.137}
\newcommand{\urst}{0.013}

\newcommand{\ac}{1.393}
\newcommand{\uac}{0.014}
\newcommand{\tdurb}{3.5722}
\newcommand{\utdurb}{0.0030}
\newcommand{\tdurd}{4.288}
\newcommand{\utdurd}{0.039}
\newcommand{\tdure}{1.899}
\newcommand{\utdure}{0.013}

\newcommand{\loggst}{4.3437}
\newcommand{\uloggst}{0.0063}
\newcommand{\fe}{0.38}
\newcommand{\ufe}{0.05}
\newcommand{\teff}{5552}
\newcommand{\uteff}{75}

\newcommand{\kms}{\ensuremath{\rm km\,s^{-1}}}
\newcommand{\ms}{\ensuremath{\rm m\,s^{-1}}}
\newcommand{\mssq}{\ensuremath{\rm m\,s^{-2}}}
\newcommand{\gcc}{\ensuremath{\rm g\,cm^{-3}}}

\usepackage{xcolor}
\usepackage{hyperref}
\usepackage{breakurl}
\usepackage{amsmath}
\usepackage{graphicx}
\usepackage{verbatim}
\usepackage{booktabs}

\slugcomment{}

\shorttitle{WASP-47 Masses}
\shortauthors{Vanderburg et al.}

\begin{document}


\title{Precise Masses in the WASP-47 System}
\author{Andrew Vanderburg\altaffilmark{18,1,$\dagger$,$\star$},
Juliette C. Becker\altaffilmark{2,$\dagger$},
Lars A. Buchhave\altaffilmark{3},
Annelies Mortier\altaffilmark{4},
Eric Lopez\altaffilmark{5},  
Luca Malavolta\altaffilmark{6,17},
Rapha\"{e}lle D. Haywood\altaffilmark{1,$\star$},
David W. Latham\altaffilmark{1},
David Charbonneau\altaffilmark{1},
Mercedes L\'opez-Morales\altaffilmark{1},
Fred C. Adams\altaffilmark{2,14},
Aldo Stefano Bonomo\altaffilmark{8},
Fran\c{c}ois Bouchy\altaffilmark{7},
Andrew Collier Cameron\altaffilmark{4}, 
Rosario Cosentino\altaffilmark{9},
Luca Di Fabrizio\altaffilmark{9},
Xavier Dumusque\altaffilmark{7},
Aldo Fiorenzano\altaffilmark{9},
Avet Harutyunyan\altaffilmark{9},
John Asher Johnson\altaffilmark{1},
Vania Lorenzi\altaffilmark{9,15,16},
Christophe Lovis\altaffilmark{7}, 
Michel Mayor\altaffilmark{7},
Giusi Micela\altaffilmark{10},
Emilio Molinari\altaffilmark{9,13},
Marco Pedani\altaffilmark{9},
Francesco Pepe\altaffilmark{7}, 
Giampaolo Piotto\altaffilmark{6,17},
David Phillips\altaffilmark{1},
Ken Rice\altaffilmark{11},
Dimitar Sasselov\altaffilmark{1},
Damien S\'egransan\altaffilmark{7},
Alessandro Sozzetti\altaffilmark{8},
St\'ephane Udry\altaffilmark{7},
Chris Watson\altaffilmark{12}
}


\altaffiltext{1}{Harvard--Smithsonian Center for Astrophysics, 60 Garden St., Cambridge, MA 02138, USA}
\altaffiltext{2}{Astronomy Department, University of Michigan, Ann Arbor, MI 48109, USA}
\altaffiltext{3}{Centre for Star and Planet Formation, Natural History Museum of Denmark \& Niels Bohr Institute, University of Copenhagen, \O ster Voldgade 5-7, DK-1350 Copenhagen K.}
\altaffiltext{4}{Centre for Exoplanet Science, SUPA, School of Physics \& Astronomy, University of St Andrews, St Andrews KY16\,9SS,\,UK}
\altaffiltext{5}{NASA Goddard Space Flight Center, Greenbelt, MD 20771, USA}
\altaffiltext{6}{Dipartimento di Fisica e Astronomia ``Galileo Galilei'', Universita'di Padova, Vicolo dell'Osservatorio 3, 35122 Padova, Italy}
\altaffiltext{7}{Observatoire Astronomique de l'Universit\'e de Gen\`eve, 51 chemin des Maillettes, 1290 Versoix, Switzerland}
\altaffiltext{8}{INAF -- Osservatorio Astrofisico di Torino, Via Osservatorio 20, I-10025 Pino Torinese, Italy}
\altaffiltext{9}{INAF -- Fundaci\'on Galileo Galilei, Rambla Jos\'e Ana Fern\'andez P\'erez, 7, 38712 Bre\~na Baja, Spain}
\altaffiltext{10}{INAF -- Osservatorio Astronomico di Palermo, Piazza del Parlamento 1, 90124 Palermo, Italy}
\altaffiltext{11}{SUPA, Institute for Astronomy, Royal Observatory, University of Edinburgh, Blackford Hill, Edinburgh EH93HJ, UK}
\altaffiltext{12}{Astrophysics Research Centre, School of Mathematics and Physics, Queen’s University Belfast, Belfast BT7 1NN, UK}
\altaffiltext{13}{INAF -- IASF Milano, via Bassini 15, 20133, Milano, Italy}
\altaffiltext{14}{Physics Department, University of Michigan, Ann Arbor, MI 48109, USA}
\altaffiltext{15}{Instituto de Astrof\'isica de Canarias, C/ V\'a L\'actea, s/n, 38205 - La Laguna (Tenerife), Spain}
\altaffiltext{16}{Departamento de Astrof\'isica, Universidad de La Laguna, Tenerife, Spain}
\altaffiltext{17}{INAF - Osservatorio Astronomico di Padova, Vicolo dell'Osservatorio 5, 35122 Padova, Italy}
\altaffiltext{18}{Department of Astronomy, The University of Texas at Austin, Austin, TX 78712, USA}
\altaffiltext{$\dagger$}{NSF Graduate Research Fellow}
\altaffiltext{$\star$}{NASA Sagan Fellow}


\begin{abstract}
We present precise radial velocity observations of WASP-47, a star known to host a hot Jupiter, a distant Jovian companion, and, uniquely, two additional transiting planets in short-period orbits: a super-Earth in a $\approx$ 19 hour orbit, and a Neptune in a $\approx$ 9 day orbit. We analyze our observations from the HARPS-N spectrograph along with previously published data to measure the most precise planet masses yet for this system. When combined with new stellar parameters and reanalyzed transit photometry, our mass measurements place strong constraints on the compositions of the two small planets. We find unlike most other ultra-short-period planets, the inner planet, WASP-47 e, has a mass (\mple\ $\pm$ \umple\ \mearth) and radius (\rple\ $\pm$ \urple\ \rearth) inconsistent with an Earth-like composition. Instead, WASP-47 e likely has a volatile-rich envelope surrounding an Earth-like core and mantle. We also perform a dynamical analysis to constrain the orbital inclination of WASP-47 c, the outer Jovian planet. This planet likely orbits close to the plane of the inner three planets, suggesting a quiet dynamical history for the system. Our dynamical constraints also imply that WASP-47 c is much more likely to transit than a geometric calculation would suggest. We calculate a transit probability for WASP-47 c of about 10\%, more than an order of magnitude larger than the geometric transit probability of 0.6\%.

\end{abstract}

\keywords{ planets and satellites: detection,  planets and satellites: gaseous planets}

\section{Introduction}

Among the many scientific results from the K2 mission, the discovery of two additional transiting planets in the known hot Jupiter WASP-47 system was perhaps the most surprising. After the second of the \Kepler\ space telescope's four reaction wheels failed and the spacecraft was re-purposed to undertake the K2 mission \citep{howell}, the parade of stunning results from the original \Kepler\ mission --- from low-density multi-transiting systems \citep{kepler11}, to evaporating and disintegrating sub-Mercury-sized planets \citep{rappaport}, to an abundance of small, likely rocky planets in their host stars' habitable zones \citep{dressingcharbonneau, petigura2, burke} --- was largely expected to slow. The \Kepler\ era of discovery was instead expected to give way to a new era in which the K2 mission would largely find more of the same kinds of planets that had been previously discovered, just orbiting brighter stars or in different environments. While K2 has certainly delivered in that regard, finding small planets around nearby stars well-suited to detailed characterization \citep{hip116454, sinukoffmultis, hip41378, rodriguez, crossfield106}, temperate planets around M-dwarfs \citep{crossfield, petiguratwoearthsized, martinez, dressingk2stars, dressingk2-2}, and planets in open clusters \citep{zeit1, zeit3, david, obermeier, libralato, zeit4},  K2 is still delivering new and unexpected discoveries, like the unique architecture of the \thisstar\ system, that are yielding fundamental insights about the formation and evolution of planetary systems. 

The \thisstar\ planetary system was first discovered by the ground-based Wide Angle Search for Planets (WASP) survey \citep{hellier}. After detecting a candidate hot Jupiter in a 4.16 day orbital period with the WASP-South instrument, \citet{hellier} followed-up the system and confirmed the planetary nature of \thisfirstplanet\ with a transit observation with the EulerCam photometer and moderate precision radial velocity observations from the CORALIE spectrograph, both on the 1.2m Euler telescope at La Silla Observatory in Chile. Several years later, \thisstar\ happened to lie in Field 3 of the K2 mission, and was observed by K2 between November 2014 and February 2015. In addition to the previously known hot Jupiter, the precise K2 photometry revealed two other transiting planets, an interior super-Earth in a 19 hour orbit, and an exterior Neptune-sized planet in a 9 day orbit, making \thisstar\ the first and only hot Jupiter system with additional short-period transiting planets \citep{becker}. Meanwhile, long-term radial velocity monitoring of \thisstar\ with CORALIE was also revealing interesting trends. Using 48 observations obtained over the course of almost 5 years, \citet{neveuvanmalle} detected another giant planet orbiting \thisstar\ in a nearly 600-day orbit, giving a total of four known planets around \thisstar. 

Although in 2015, when \thisstar\ c, d, and e were discovered, extra transiting planets in a hot Jupiter system seemed unusual and surprising, such planets were once believed likely to exist, and were in fact seen as a highly promising way to find small transiting planets before the launch of multi-million-dollar wide-field space telescopes like \Kepler\ and CoRoT. \citet{holmanmurray} and \citet{agol} showed that a transiting planet would undergo small deviations from perfectly periodic transits (called transit timing variations or TTVs) in the presence of other nearby planets in the system, and \citet{steffenagol} showed that this method is highly sensitive to small planets orbiting near hot Jupiters. Frustratingly, however, after a decade of searching, the TTV method's exquisite sensitivity to small planets near hot Jupiters had merely translated to exquisite upper limits on the presence of such planets \citep[especially in mean motion resonances with the hot Jupiter,][]{steffenagol, millerricci, collins}. 

Meanwhile, shortly after its launch, the \Kepler\ telescope detected the first transit timing variations in systems of longer period and lower mass planets than the hot Jupiters on which previous searches had focused \citep{kepler9, kepler11}. Over the course of its mission, \Kepler\ found planets in nearly every configuration imaginable\footnote{Some exceptions to this statement include the lack of binary planets, Trojan planets,  and circumtrinary planets among \Kepler's discoveries.}, including tightly packed systems of small planets with short orbital periods \citep{muirhead} and multi-planet systems with slightly longer period warm Jupiters \citep{bonomo, sanchisojedausp, huangwarmjupiter}, but \Kepler\ found no evidence for any additional planets near a hot Jupiter. When detailed investigations and searches for companions to hot Jupiters in \Kepler\ data came up empty \citep{steffenconstraints}, the scientific community largely considered the issue resolved --- hot Jupiters evidently either cannot or almost never have nearby planetary companions.

Therefore, the planets around \thisstar\ must represent a rare outcome of planet formation, and any observational or theoretical insights into their architecture and origins are important to help illuminate this new mode. Follow-up work came quickly. \citet{sanchisojedaw47rm} detected the Rossiter McLaughlin effect for \thisfirstplanet, ruling out large misalignments between the inner transiting system's orbits and the star's sky-projected spin axis. Radial velocity monitoring with the Planet Finder Spectrograph (PFS) on the Magellan Clay telescope detected the reflex motion due to \thissecondplanet\ and found that its composition was most likely rocky \citep{dai}. More recently, a larger set of precise velocities from the HIgh Resolution Echelle Spectrometer (HIRES) on the Keck I telescope obtained by \citet{sinukoffw47} improved the precision on \thissecondplanet's mass and detected \thisthirdplanet's RV signature. A photodynamical analysis by \citet{almenara} and a simultaneous analysis of radial velocities and transit times by \citet{weissw47} placed even stronger constraints on the planets' masses and eccentricities, showing that \thisthirdplanet\ had a mass close to that of Neptune, and that the eccentricities of the inner planets were small. \citet{beckeradams} used the fact that the three inner planets all transit to place constraints on the inclination of \thisfourthplanet. They showed that \thisfourthplanet's orbit is probably fairly well aligned with the transiting planets, though certain highly misaligned orbits could still allow the inner planets to transit.

Meanwhile, others have speculated about the origin of the \thisstar\ planets. \citet{batygin} suggested that {\em in situ} formation could be an important channel for creating hot Jupiters, and that small planetary companions like those around \thisstar\ (or in orbits misaligned with the hot Jupiter) would be a natural consequence of this mechanism. \citet{huangwarmjupiter} also suggested {\em in situ} formation, noting that the planets around \thisstar\ are much more reminiscent of planetary systems hosting warm Jupiters than other hot Jupiters, and speculating that \thisstar\ might be an extreme short-period result of an {\em in situ} warm Jupiter formation mechanism. On the other hand, \citet{weissw47} suggested that \thisstar's planets might have formed in a two-stage process, where the two Jovian planets formed far out in the disk, \thisfirstplanet\ migrated inwards, and then the two smaller planets  formed nearby. More constraints on the {\ron planets'} masses and compositions and the system architecture are needed to understand how these unusual planets formed and came to be in their present configuration. 

In this paper, we add to the already large body of follow-up work on the \thisstar\ system, presenting {\ron69} new precise radial velocity observations from the High Accuracy Radial velocity Planet Searcher for the Northern hemisphere (HARPS-N). We analyze these new observations along with previously collected data to determine the most precise values yet for the masses and radii of the \thisstar\ planets. In Section \ref{observations}, we describe our HARPS-N dataset. Section \ref{analysis} describes our analysis of the HARPS-N data to measure spectroscopic properties and planet masses, as well as a re-analysis of the K2 light curve, and a re-determination of the host star's physical properties. In Section \ref{dynamics}, we use N-body integrations to simulate the \thisstar\ system using our new measurements of the planetary masses, and place constraints on the inclination of the long-period, not necessarily transiting planet \thisfourthplanet. Finally, in Section \ref{discussion}, we discuss what our measurements tell us about the compositions of the \thisstar\ planets and what our dynamical constraints tell us about the system's history.



\section{HARPS-N Spectroscopy}\label{observations}

We observed \thisstar\ with the HARPS-N spectrograph on the 3.58m Telescopio Nazionale Galileo (TNG) on the island of La Palma, Spain \citep{harpsn}. HARPS-N is a stabilized high resolution ($\lambda/\Delta\lambda = 115,000$) optical spectrograph designed specifically to make precise radial velocity measurements. We began observing \thisstar\  on 23 July 2015, shortly after K2 data revealed the presence of two small transiting planets (\thissecondplanet\ and \thisthirdplanet) in addition to the previously known hot Jupiter (\thisfirstplanet). We obtained 78 observations of \thisstar\ with integration times of 30 minutes. We measured radial velocities for each exposure by cross-correlating the HARPS-N spectra with a weighted binary mask \citep{baranne, pepe}.  The 30 minute exposures of \thisstar\ yielded radial velocity measurements with typical photon-limited uncertainties of 3 \ms. HARPS-N is fed by two optical fibers going into the spectrograph - one fiber feeds the target starlight into the spectrograph, while the other fiber either feeds in a precise wavelength calibrator\footnote{The wavelength calibrator light source can be a Thorium Argon lamp, continuum light passed through a stabilized Fabry-Perot interferometer, or a stabilized laser frequency comb.}, or sky background light. HARPS-N is stable to better than 1 \ms\ over the course of a night, considerably more precise than our typical measurement uncertainties, so a simultaneous calibrator was not necessary for our observations. Instead, we fed sky background light into the instrument with the second fiber. 

Because \thisstar\ is somewhat faint for precise radial velocity measurements, and because it lies in the ecliptic plane near bright solar system objects like the moon, contamination from scattered sky background light can cause significant velocity errors in some conditions. We identified radial velocity observations contaminated by scattered light from a bright sky using the method described by \citet{malavolta}. In brief, we calculated radial velocities with and without sky contribution removed (using the sky spectrum obtained simultaneously with the instrument's second fiber), and flagged the exposures that showed a significant (2--$\sigma$) velocity difference with and without sky subtraction. We found that four of our 78 observations showed evidence for sky contamination and excluded them from our analysis (which we describe in Section \ref{sec:rv_analysis}).

\section{Analysis}\label{analysis}

\subsection{Spectroscopic Parameters}
\label{sec:spec_prop}
We used our HARPS-N spectra to measure spectroscopic parameters for \thisstar. We first used the Stellar Parameter Classification (SPC) method \citep[][]{buchhave, buchhave14}. SPC works by cross-correlating a stellar spectrum with synthetic spectra from \citet{kurucz} model atmospheres and interpolating the resulting correlation peaks to determine stellar atmospheric parameters like effective temperature, surface gravity, metallicity, and line broadening. We ran SPC on 75 of the 78 HARPS-N spectra\footnote{We excluded several spectra due to their low signal-to-noise ratios. {\ron We did not exclude the sky-contaminated spectra from our SPC analysis because the small amount of sky contamination necessary to skew the radial velocity by $\approx$ 10 \ms\ does not significantly affect the SPC analysis.}} and calculated the averages for all of the spectroscopic parameters. With SPC, we measure a temperature of $T_{\rm eff, SPC }=$ 5571 $\pm$ 50 K, a surface gravity of $\log{g_{\rm cgs, SPC}}= $ 4.39 $\pm$ 0.1, a metallicity [M/H]$_{\rm SPC}$ = 0.42 $\pm$ 0.08, and place an upper limit on rotational velocity of $v_{\rm rot}$ \textless\ 2 \kms. The error bars reported for SPC reflect systematic uncertainties in the stellar models used by SPC. The scatter in the parameters SPC reports for each of the 75 individual spectra is much smaller than the systematic uncertainty in these parameters.   

We also measured spectroscopic parameters using the method described by \citet{mortier} on our new HARPS-N spectra. We co-added all of the HARPS-N spectra, measured equivalent widths of iron lines using ARES2 \citep{sousa, ares2}, and determined atmospheric parameters using the 2014 version of MOOG\footnote{\url{http://www.as.utexas.edu/~chris/moog.html}} \citep{sneden1973}. We then applied the surface gravity correction from \citet{mortier14} to adjust for systematic effects in the ARES/MOOG analysis method. This analysis yielded an effective temperature $T_{\rm eff, MOOG =}$ 5614 $\pm$ 67 K, surface gravity $\log{g_{\rm cgs, MOOG}}= $ 4.45 $\pm$ 0.11, and an iron abundance [Fe/H]$_{\rm MOOG}$ = 0.39 $\pm$ 0.05.

The spectroscopic parameters that we determined through our SPC and ARES/MOOG analyses are consistent with one another, and are also consistent within errors with previous determinations. In this paper, we adopt weighted averages of the spectroscopic parameters from our SPC analysis and our ARES/MOOG analysis, along with the spectroscopic analysis done  by \citet{sinukoffw47} on a high signal-to-noise spectrum from Keck Observatory using another independent method, Spectroscopy Made Easy \citep[SME, ][]{brewer2015, brewer2016}. Each of the three methods we use have their own systematics, so averaging the three methods could give more accurate spectroscopic parameters. The weighted averages were a temperature of $T_{\rm eff} =$ 5552 K, a metallicity\footnote{To calculate the weighted average metallicity, we use iron abundance as a proxy for the overall metallicity {\ron by assuming solar abundance ratios, where [M/H] = [Fe/H].}} of [M/H] = 0.38, and a stellar surface gravity of $\log{g_{\rm cgs}} =$ 4.32. We assigned conservative error bars of 75 K in temperature, 0.05 dex in metallicity, and 0.1 dex in surface gravity to the weighted averages in order to account for systematic errors in the spectroscopic analyses.

\subsection{Transit Re-analysis}
\label{sec:lc_analysis}

Since the original analysis of the \thisstar\ K2 light curve by \citet{becker}, there have been new data collected \citep{sinukoffw47} and additional dynamical analyses \citep{almenara, weissw47} performed, which put tighter constraints on the planets' orbits (in particular, their eccentricities). In this section, we re-analyze the K2 light curve using Kepler's third law to link together the planets' scaled semi-major axes while taking into account new, tighter constraints on the inner planets' orbital eccentricities. These constraints yield better measurements of the planets' scaled semi-major axes, and therefore better measurements of the planets' transit impact parameters and planet to star radius ratios. 

We performed our new transit analysis using the same K2 light curve of \thisstar\ produced and used by \citet{becker}.  \Kepler\ observed \thisstar\ for 69 days between 2014 November 14 and 2015 January 23 in short-cadence mode -- an exposure of \thisstar\ was recorded every 58.85 seconds instead of \Kepler's normal 29.4 minute ``long cadence'' integrations. After \citet{becker} processed the light curve to remove systematic effects due to \Kepler's unstable pointing \citep[see also ][]{vj14,v15},  the photometric scatter in the light curve was about 350 ppm per minute. 

We perform the transit analysis on the K2 light curve using a Markov Chain Monte Carlo (MCMC) algorithm with an affine invariant ensemble sampler \citep{goodman}. We fit the three transiting planet light curves with \citet{mandelagol} model light curves (with a quadratic limb darkening law parameterized following \citealt{kippingld}).  We fit for the planets' orbital periods, transit ephemerides\footnote{Even though small transit timing variations have been detected in the K2 light curve, for this analysis, we assumed the transits of the \thisstar\ planets are perfectly periodic. We have also analyzed the K2 light curve while shifting the transit center times to account for the transit timing variations, and found the difference in fitted parameters was negligible.}, planet to star {\ron radius} ratios ($R_p/R_\star$), orbital inclinations ($i$), and in some cases, the orbital eccentricity ($e$) and argument of periastron ($\omega_p$). Instead of fitting for all three planets' scaled semi-major axis ratios ($a/R_\star$) independently, we fit for stellar density and calculated $a/R_\star$ for each planet using Kepler's third law. We also fit for a single flux offset parameter and the uncertainty of each K2 photometric datapoint. 

We force the orbits of \thissecondplanet\ and \thisfirstplanet\ to be circular; the tidal circularization timescales for these two planets ($10^5 $ years and $10^7$ years respectively, using the expression from \citealt{goldreich} and reasonable values\footnote{In particular, we used $Q/k_2 = 10^2$ for {\ron \thissecondplanet} and $Q/k_2 = 10^5$ for \thisfirstplanet\ and \thisthirdplanet.} of $Q$ and $k_2$) are much shorter than the age of the system{\ron\footnote{\ron Planet-planet interactions will drive a small forced eccentricity but our dynamical calculations show that typical forced eccentricities for these planets are of order $10^{-3}$, too small to affect our measured parameters.}}. For \thisthirdplanet, while tidal dissipation is not strong enough to necessarily circularize the orbit, strong dynamical constraints exist on any eccentricity. Both N-body simulations performed by \citet{becker} and simultaneous analysis of transit times and radial velocities by \citet{almenara} and \citet{weissw47} have showed that the eccentricity must be quite small. We imposed a half-Gaussian prior on eccentricity centered at 0 with a 1-$\sigma$ width of 0.014, and solutions with eccentricity less than 0 forbidden. This prior matches the distribution of dynamically stable simulations from \citet{becker} and gives a 2-$\sigma$ upper limit on eccentricity that matches the limit from \citet{weissw47}. In our MCMC fit, we explored eccentric models for \thisthirdplanet's orbit by stepping in $\sqrt{e}\sin{\omega_p}$ and $\sqrt{e}\cos{\omega_p}$. 

We initialized an ensemble of 100 walkers, evolved them for 20,000 steps, and removed the first 10,000 steps, when the ensemble was ``burning in'' and not yet in a converged configuration. We assessed convergence of the resulting MCMC chains by calculating the Gelman Rubin statistic, which was less than 1.2 for all parameters, and less than 1.05 for most. The best-fit parameters and their uncertainties are reported in Table \ref{bigtable}. The results are consistent with the previous transit analyses by \citet{becker} and \citet{almenara}, but in some cases are more precise because of the additional constraints we have imposed here. Thanks to the high signal-to-noise transits of \thisfirstplanet, we measure the stellar limb darkening coefficients\footnote{We measure the linear coefficient $u_1 = 0.533 \pm 0.010$ and the quadratic coefficient $u_2 = 0.097 \pm 0.024$. These coefficients are in in reasonable agreement (within $\approx$ 0.1) with the theoretical predictions from \citet{claretbloemen} of $u_1 = 0.47$ and $u_2 = 0.22$.} and importantly, we measure the stellar density with a precision of 1.4\%, which we use in Section \ref{sec:stellar_prop} to determine a precise stellar mass and radius. We also measure the radius ratios of \thissecondplanet\ and \thisthirdplanet\ with 0.8\% and 0.5\% precision, respectively.

\subsection{Stellar Parameters}
\label{sec:stellar_prop}

In this section, we take advantage of the well measured stellar density we measure from the K2 transit photometry to derive precise stellar parameters \citep[as has been done previously for other hot Jupiter hosts, e.g.][]{sozzetti07}. We base our analysis on the Yonsei-Yale (YY) isochrones \citep{Yi:2001}, exploring parameter space stepping in stellar mass, age, and metallicity using MCMC with an affine invariant ensemble sampler. We interpolated the YY isochrones using code\footnote{Code available at \url{https://github.com/jdeast/EXOFASTv2/}.} written by Jason Eastman for EXOFASTv2 (\citealt{exofast, rodriguez}; Eastman et al. in prep).  We impose Gaussian priors on \thisstar's density, metallicity, effective temperature, and surface gravity. The density prior's center and width come from our analysis of the K2 light curve. For the temperature, metallicity, and surface gravity priors, we use the values and uncertainties from the weighted average of the three different spectroscopic analyses we discussed in Section \ref{sec:spec_prop}. 

Thanks to our precise measurement of \thisstar's density from the transit light curves, our MCMC analysis yielded a stellar mass and radius with precisions of about 3\% and 1\% respectively. Stellar evolutionary models have not yet been tested at such high precisions on stellar parameters \citep{torresmasses}, so we performed tests to assess the scale of systematic errors in the isochrones we used.  First, we compared the results of our analysis with the YY isochrones to the similar analysis performed by \citet{almenara} using the Dartmouth isochrones \citep{Dotter:2008}. Using the same priors and input stellar parameters, our YY analysis yielded a mass about 1.5\% larger and a radius about 0.5\% larger than the values  \citet{almenara} determined using the Dartmouth isochrones. The differences in mass and radius between the YY analysis and Dartmouth analysis are about half the size of the uncertainties from each analysis. We also performed an MCMC analysis to determine \thisstar's stellar parameters using the empirical mass and radius relations determined by \citet{torresmasses}. We stepped in surface gravity, metallicity, and effective temperature, imposing Gaussian priors on the metallicity, temperature, and stellar density. The Torres relations have a known offset {\ron at} one solar mass, over-predicting the mass of the Sun by about 5\% and the radius of the Sun by about 2\% \citep{torresmasses}, so we scaled the output masses and radii inside our model by those factors to correct for the offset -- forcing the relations to correctly predict the Sun's density. Our analysis using the Torres relations yielded masses and radii that were about 0.7\% and 0.1\% larger than our analysis using the YY isochrones. These discrepancies again are considerably smaller than the uncertainties we determined in our stellar parameters. Finally, we repeated the analysis with the Torres relations while applying the empirical correction for solar-mass stars provided by \citet{santos}, and found that it also gave consistent values for the stellar mass and radius, with values about 0.5--$\sigma$ larger than the YY value. 

Based on these tests showing different models and relations all predicting consistent stellar parameters for \thisstar, we conclude that systematic uncertainties in the stellar masses and radii we derive are small. This makes sense -- stellar evolutionary models are calibrated off the Sun and incorporate physics known to be important for Sun-like stars, and therefore tend to be most accurate for stars with parameters close to those of the Sun. \thisstar\ has a mass only 4\% larger than the sun, and is only slightly more evolved, with a $\approx$ 14\% larger radius and $\approx$ 200 K cooler stellar effective temperature. The biggest discrepancy between \thisstar\ and the Sun is the composition -- \thisstar\ has a metallicity 2.5 times higher than that of the Sun, which is possibly the cause of the small discrepancies we do see between the different methods.  

To take the 0.5--$\sigma$ discrepancies we found into account, we re-determined the stellar mass and radius using our YY MCMC analysis after inflating the uncertainty on stellar density by adding an 0.5--$\sigma$ systematic uncertainty in quadrature to the uncertainty we derived from our transit analysis. This analysis yielded a stellar mass of \mst\ $\pm$ \umst\ \msun\ and a radius of \rst\ $\pm$ \urst\ \rsun. Our constraint on stellar density is precise enough that we also measure an isochronal age of 6.7$^{+1.5}_{-1.1}$ Gyr for \thisstar, although the precision of this age determination also pushes the level at which isochronal ages are accurate \citep[isochronal ages can have systematic errors of up to 25\%, ][]{torresmasses}.

\subsection{Radial Velocity Analysis}
\label{sec:rv_analysis}

\begin{figure}[ht!] 
   \centering
   \includegraphics[scale=0.6]{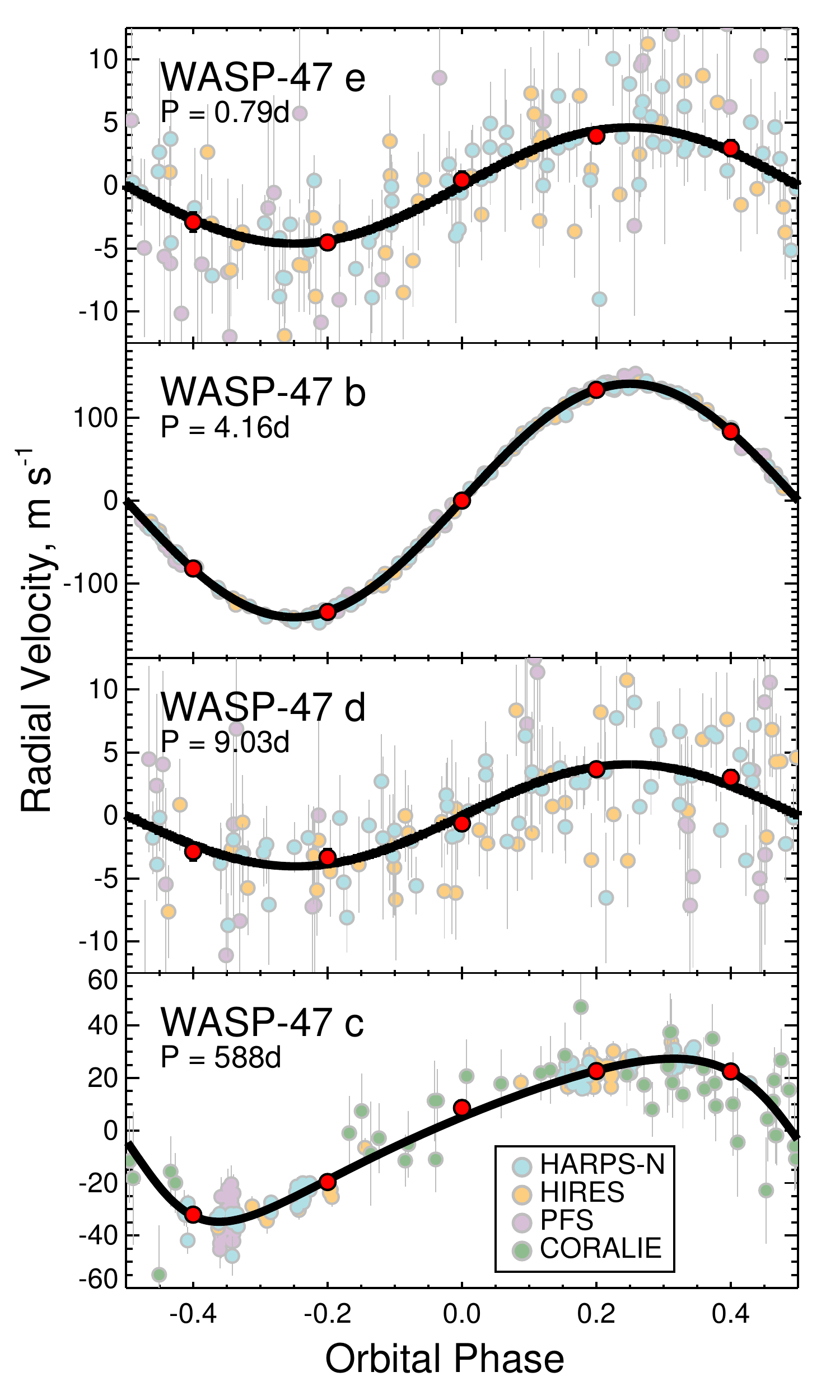} 
   \caption{Radial velocity observations of \thisstar\ from four spectrographs, folded on the periods of the four planets, with the best-fit model for each of the other three planets subtracted away. Data from HARPS-N are shown as pale blue dots, HIRES as pale orange dots, PFS as pale maroon dots, and CORALIE as pale green dots. The CORALIE data are only shown for \thisfourthplanet, as they are not of high enough precision to meaningfully constrain the orbits of \thissecondplanet\ and \thisthirdplanet. The thick black lines are the best-fit models for each of the four planets, and the dark red dots are binned points. We note that we do not include the PFS velocities in our final analysis.}
   \label{rvphase}
\end{figure}

\begin{deluxetable}{lrc}
\tablewidth{0pt}
\tablecaption{ HARPS-N Radial Velocities of \thisstar\ \label{rvs}}
\tablehead{
\colhead{BJD - 2454833} & \colhead{RV [\ms]} &\colhead{$\sigma_{\rm RV}$ [\ms]} }
\startdata
2457226.561 & -27014.60 & 6.43\\
2457226.699 & -26987.90 & 4.84\\
2457226.721 & -26982.21 & 6.71\\
2457227.561 & -26937.25 & 2.63\\
2457227.713 & -26954.63 & 2.55\\
2457228.566 & -27113.16 & 4.71\\
2457228.700 & -27142.69 & 4.98\\
2457229.562 & -27219.22 & 5.11\\
2457229.703 & -27190.29 & 3.24\\
2457230.572 & -27039.26 & 2.59\\
2457230.711 & -27010.20 & 2.34\\
2457254.520 & -27210.88 & 3.77\\
2457256.548 & -26932.08 & 2.20\\
2457256.641 & -26933.99 & 3.06\\
2457257.644 & -27101.64 & 3.34\\
2457267.487 & -27151.78 & 2.43\\
2457267.602 & -27127.38 & 2.23\\
2457268.521 & -26968.72 & 7.43\\
2457268.624 & -26945.85 & 2.77\\
2457269.542 & -26995.84 & 2.31\\
2457270.526 & -27188.86 & 1.85\\
2457271.530 & -27172.97 & 1.88\\
2457272.568 & -26967.95 & 2.45\\
2457273.544 & -26970.22 & 2.09\\
2457301.412 & -27017.29 & 1.95\\
2457301.521 & -26992.78 & 2.21\\
2457302.411 & -26931.79 & 2.08\\
2457302.524 & -26936.49 & 2.99\\
2457324.337 & -27134.85 & 3.00\\
2457325.323 & -27192.62 & 3.33\\
2457325.417 & -27190.58 & 2.67\\
2457330.315 & -27046.38 & 2.67\\
2457330.443 & -27031.60 & 2.93\\
2457331.316 & -26930.42 & 2.97\\
2457331.421 & -26929.30 & 2.55\\
2457333.328 & -27202.65 & 2.60\\
2457336.316 & -27018.83 & 2.47\\
2457336.398 & -27036.13 & 2.99\\
2457557.701 & -27149.24 & 2.56\\
2457558.695 & -27093.68 & 2.09\\
2457559.702 & -26902.75 & 2.53\\
2457560.684 & -26931.92 & 3.67\\
2457562.695 & -27116.92 & 3.47\\
2457563.687 & -26918.20 & 4.10\\
2457565.691 & -27101.52 & 3.66\\
2457566.698 & -27143.80 & 2.79\\
2457573.722 & -27054.63 & 3.49\\
2457574.613 & -27167.37 & 2.72\\
2457574.701 & -27157.09 & 2.51\\
2457576.584 & -26880.69 & 3.16\\
2457576.685 & -26886.91 & 2.42\\
2457579.695 & -27041.15 & 7.68\\
2457580.680 & -26877.83 & 7.44\\
2457616.560 & -27138.59 & 6.88\\
2457617.581 & -26939.47 & 1.98\\
2457618.579 & -26882.95 & 2.53\\
2457651.460 & -26875.57 & 2.16\\
2457652.459 & -26982.58 & 2.18\\
2457653.463 & -27155.02 & 2.81\\
2457654.464 & -27057.53 & 2.88\\
2457655.492 & -26870.08 & 5.08\\
2457658.510 & -27083.94 & 4.73\\
2457659.516 & -26895.47 & 2.20\\
2457661.490 & -27123.19 & 2.53\\
2457669.458 & -27063.11 & 2.80\\
2457670.456 & -27140.53 & 3.07\\
2457671.451 & -26981.35 & 1.97\\
2457672.454 & -26870.44 & 1.88\\
2457721.368 & -27000.32 & 2.36\\

\enddata
\label{rvtable}
\end{deluxetable}

\begin{deluxetable*}{lcccc}[h!]
\tablecaption{Summary of Radial Velocity Observations of \thisstar \label{rvsummarytable}}
\tablewidth{0pt}
\tablehead{
\colhead{} & \colhead{HARPS-N} & \colhead{HIRES}& \colhead{PFS$^{a}$ }& \colhead{CORALIE}}
\startdata
Number of usable observations &  69  & 43 & 26 & 46 \\  
Standard deviation about best-fit & 3.3 \ms & 3.7 \ms & 7.4 \ms  & 13.5 \ms\\  
Mean photon-limited uncertainty & 3.3 \ms & 2.0 \ms  & 3.2 \ms &12.1 \ms \\
Time baseline & 495 days & 412 days  & 12 days & 1622 days
\enddata
\tablecomments{$a$: In our final analysis, we exclude the PFS data. The standard deviation reported in the table is about the best-fit solution which did not include the PFS data. When the PFS data is included in the fit, its standard deviation about the best-fit model is 7.3 \ms.} 
\label{tbl:rvdatasets}
\end{deluxetable*}

We analyzed our radial velocity observations and archival observations taken from the literature to measure masses and orbital parameters for the \thisstar\ planets. In our analysis, we combined our HARPS-N observations with previously published radial velocities from CORALIE \citep{hellier, neveuvanmalle}, PFS \citep[][]{dai}, and HIRES \citep{sinukoffw47}. From our HARPS-N dataset, we excluded the four points we found to be contaminated by sky background light and one point that was taken during poor conditions and which exhibited a photon-limited velocity uncertainty of 16 \ms, six times greater than the typical uncertainty in our dataset. From the HIRES dataset, we excluded the points taken on the night of their Rossiter-McLaughlin observation \citep{sanchisojedaw47rm, sinukoffw47}. From all four RV datasets, we excluded points taken within two hours of the mid-transit time of \thisfirstplanet\ since those points are affected by the planet's Rossiter-McLaughlin signal which we do not model\footnote{This led to the exclusion of four HARPS-N data points and two HIRES datapoints.}. The Rossiter-McLaughlin effects of the two other transiting planets are negligibly small so we retained points taken during their transits. We list our HARPS-N velocity observations in Table \ref{rvtable} and summarize the four datasets in Table \ref{tbl:rvdatasets}. 

We modeled the radial velocity of \thisstar\ as a sum of four Keplerian functions, and do not attempt to model the gravitational interactions between the four planets. Even though mutual gravitational interactions do perturb the planets' orbits \citep{becker, almenara,weissw47}, the effect on the radial velocity curve is undetectably small \citep{dai, sinukoffw47}. We confirmed this result holds over the longer time-span of our observations by numerically integrating the system. We used the same eccentricity priors for the RV analysis as we did for the transit analysis -- specifically we forced the orbits of \thisfirstplanet\ and \thissecondplanet\ to be circular, and we imposed a half-Gaussian prior on the eccentricity of \thisthirdplanet\ centered at 0 (without allowing negative eccentricity solutions) with a standard deviation of 0.014. For \thisfourthplanet, we allowed eccentricity and the argument of periastron to vary freely with only uniform priors imposed. For the three transiting planets, we imposed Gaussian priors on orbital period and time of transit at the values and uncertainties we derived in our transit analysis. We also imposed a prior that the RV semiamplitudes of all four planets be greater than zero, but the signals of all four planets were detected strongly enough that this prior had no effect. 

\thisstar\ is photometrically quiet, and we see no evidence in the radial velocities or activity indicators of correlations due to stellar activity, so we used a white noise model for our RV analysis, with separate instrumental ``jitter'' terms for data from the four different telescopes. We also fit for velocity zero-point offsets for the four different instruments. We did not impose any informative priors on the jitter terms and zero-point offsets. We explored parameter space using an MCMC algorithm with affine invariant ensemble sampling, like for our transit analysis and our stellar parameter analysis. We used 100 walkers and evolved their positions for 150,000 steps each. To match the size of the chains from our transit analysis (Section \ref{sec:lc_analysis}) and our stellar parameter analysis (Section \ref{sec:stellar_prop}),  we used the final 10,000 steps in our chains to estimate parameters. We confirmed the MCMC chains were converged by calculating the Gelman-Rubin statistics; the values were below 1.05 for all parameters.   

We show the radial velocities from all four spectrographs and our best-fitting model in Figure \ref{rvphase}. From our combined analysis with data from all four spectrographs, we measure masses that are more precise than, but consistent with previous determinations, except for the mass of \thissecondplanet, which is somewhat at odds with the masses determined by \citet{dai} and \citet{sinukoffw47}. In particular, using data from all four spectrographs, we measure the mass of \thissecondplanet\ to be 7.15 $\pm$ 0.67 \mearth, about 1.5--$\sigma$ lower than both the measurements by \citet{dai} of 12.2 $\pm$ 3.7 \mearth, and \citet{sinukoffw47} of 9.11 $\pm$ 1.17 \mearth. 


We investigated the source of this discrepancy by repeating the RV fits with different datasets included and removed from the analysis. We found that the HIRES and HARPS-N datasets are quite consistent, both yielding masses for \thissecondplanet\ between 6.5 and 7 \mearth, but that the PFS dataset favors a planet mass almost a factor of two larger\footnote{The discrepancy between our combined mass measurement and that of \citet{sinukoffw47} is mostly due to the fact that \citet{sinukoffw47} included the PFS data in their analysis, which pulled their solution to higher masses.}. There are two possible explanations for the discrepancy betwen the PFS mass measurement and the HARPS-N/HIRES measurements.
\begin{enumerate}
\item The discrepancy is the result of random chance. The PFS measurement of \thissecondplanet's semi-amplitude is only 1.5--$\sigma$ away from the HARPS-N/HIRES solution, a discrepancy that should happen in about 6.5\% of all similar datasets. If this explanation is correct, then including the PFS data in our solution is appropriate and will help the mass measurement converge to the true mass.  
\item The discrepancy is the result of some time-correlated systematic errors in the PFS velocities. In this case including the data in our solution is not appropriate and will not help our measurements converge to the true mass. 
\end{enumerate}

There are reasons to believe the PFS velocities of \thisstar\ could be systematically erroneous -- the scatter of the PFS data about the solution is 7.3 \ms, more than twice the photon-limited uncertainties listed by \citet{dai}, and considerably worse than both PFS's typical RV precision \citep[better than 2 \ms,][]{teskepfs} and the radial velocity scatter from HARPS-N (3.3 \ms) and HIRES (3.7 \ms) in observations of \thisstar. Because the PFS observations were all taken over the course of only 12 days, if systematics are the cause of the large scatter in the velocities, any time correlations in the systematics would not necessarily average out.

Because of the risk of systematics contamination, and because the PFS data do not help much to constrain our velocity solution (the dataset is both smaller and less precise than the HARPS-N and HIRES observations, and the time baseline is not as long as the CORALIE observations), we choose to exclude the PFS observations from our final analysis. We re-ran the same MCMC analysis as before using only data from CORALIE, HARPS-N, and HIRES, and determined new planet masses. The masses and uncertainties of \thisstar\ b, c, and d were essentially unchanged, but the mass of \thissecondplanet\ decreased by about 0.5--$\sigma$ while the uncertainty was unchanged. The results of this RV analysis are reported in Table \ref{bigtable}. 

We measure masses of \mple\ $\pm$ \umple\ \mearth\ for \thissecondplanet, \mplb\ $\pm$ \umplb\ \mearth\ for \thisfirstplanet, \mpld\ $\pm$ \umpld\ \mearth\ for \thisthirdplanet, and a minimum mass of $M_p\sin{i} =$ \mplc\ $\pm$ \umplc\ \mearth\ for \thisfourthplanet. \thisfourthplanet's orbit is significantly eccentric, with $e_{c} =$\eccc\ $\pm$ \ueccc. The longer time-baseline of the HARPS-N observations compared to the previously published HIRES dataset and the higher precision compared to the CORALIE dataset gives a more precise measurement of the outer planet's orbital eccentricity and argument of periastron than before\footnote{We note that another possible explanation for \thisfourthplanet's eccentricity is that it is caused by the un-modeled RV signal from an additional planet with half the orbital period of \thisfourthplanet\ \citep{angladaescude}, or the un-modeled signal from a longer-period planet \citep[as postulated by][]{weissw47} contributing an RV acceleration. Our observations do not yet have the precision and time-baseline to distinguish these scenarios, so we interpret the signal as being caused by one eccentric planet. }. 


\section{Dynamical Constraints}
\label{dynamics}

Unlike the three inner planets, which have precisely known (relative) orbital inclinations from their transit light curves, \thisfourthplanet\ has only been detected in radial velocity observations so far\footnote{If \thisfourthplanet\ does happen to transit, it would not have been detected by K2 because it was at the wrong phase of its orbit during the 60 days of observations.}. In this section, we use dynamical arguments to constrain the inclination of \thisfourthplanet. We perform a variation of the analysis done by \citet{beckeradams}, who constrained \thisfourthplanet's inclination by requiring that when perturbed by the outer planet, the three inner planets in the \thisstar\ system all continually co-transit in the same orbital plane (where co-transit is defined as all of the planets crossing the star at any impact parameter, including grazing transits). \citet{beckeradams} found that this requirement does not rule out orbits of \thisfourthplanet\ with high inclinations relative to the inner three planets, but that a realization of \thisfourthplanet\ with an inclination within a degree of the central projected plane of the star will always allow the inner three planets to transit. 

\begin{figure*}[ht!] 
   \centering
   \includegraphics[width=7in]{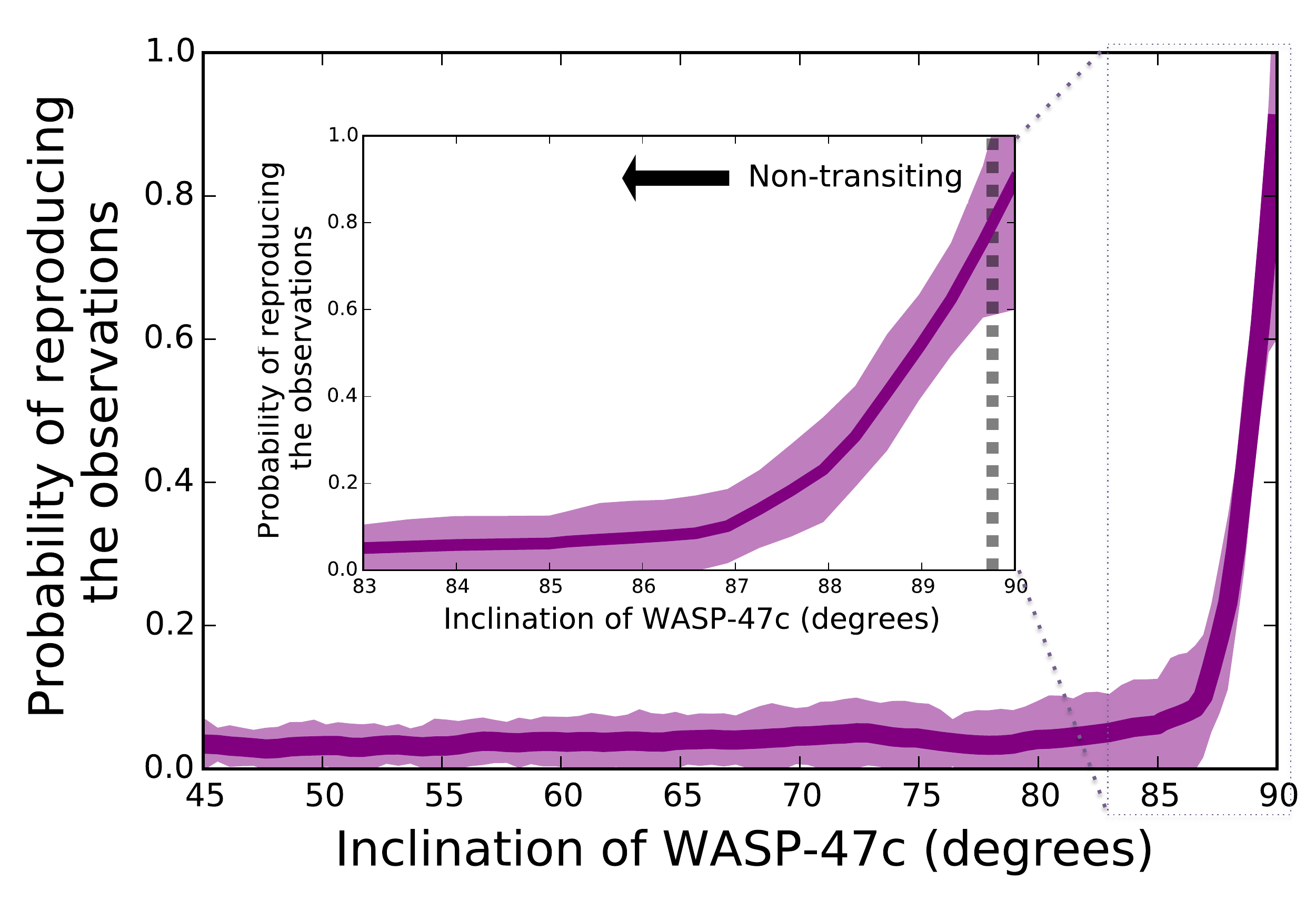} 
   \caption{Constraints on the orbital inclination of \thisfourthplanet. This plot shows the fraction of time the \thisstar\ system can reproduce our observations as a function of the inclination of \thisfourthplanet. The dark purple line is the average fraction of the time our simulations reproduce our observations of \thisstar, and the colored contours show the 1$\sigma$ range calculated from the individual realizations. The inset diagram provides a zoomed-in version of the main figure to show the detail near $i_c$  = 90$^\circ$ (i.e., near the orbital plane of the inner planets).}
   \label{inclinationlimits}
\end{figure*}
Here we significantly strengthen the constraints placed by \citet{beckeradams} by requiring that models of the \thisstar\ system with \thisfourthplanet\ orbiting at various different inclinations reproduce other observed properties of the system. We relax the constraint that the planets all transit the star in the original orbital plane, and instead  we require: 

\begin{enumerate}
\item The inner planets around \thisstar\ remain dynamically stable over integrations of 1 Myr. 
\item The inner planets have mutual inclinations such that their transit impact parameters are as close to zero as we measure (at 3$\sigma$ confidence)\footnote{Here, although the fact that all three planets transit does not technically constrain their longitudes of ascending nodes, we assume the inner three planets are indeed coplanar. If the three planets did have high mutual inclinations, the probability that all three would transit is very small -- a probability of about 1 in 1000 compared to a co-planar transit probability of roughly 1 in 15.}. 
\item \thisfirstplanet\ has a sky-projected spin/orbit obliquity consistent with the result of \citet{sanchisojedaw47rm}. 
\end{enumerate}

We conduct numerical N-body simulations using the \texttt{Mercury6} \citep{mercury6} software package to evaluate the likelihood that \thisfourthplanet\ allows these three criteria to be satisfied --- that is, the likelihood that the simulations reproduce the observations at varying values of \thisfourthplanet's orbital inclination. In our simulations, we use a hybrid symplectic and Bulirsch-Stoer (B-S) integrator, requiring energy conservation to a part in 10$^{-8}$ or better, and allowing each integration to run for 1 Myr with a starting time-step of 14 minutes. We run 2000 total 1 Myr integrations, each with randomly drawn initial conditions. 

The choice of 1 Myr as an integration time was chosen for two reasons: first, the short time-step required for these simulations is computationally demanding and integrations of 1 Myr remain feasible; second, 1 Myr encapsulates many ($>10^{8}$) dynamical times of the inner planet, and several (3-4) secular time scales. These timescales are important because orbital instabilities will occur on dynamical time scales, while motions in a single secular cycle will be expected to repeat in subsequent secular cycles. A integration time of 1 Myr allows us to effectively evaluate any dynamical instabilities, as well as encapsulate any long-term secular variations that may occur. 

The initial conditions for each planet in each integration are drawn from the observations presented in Table \ref{bigtable}. For most orbital parameters of the inner three planets, we draw from a normal distribution with mean and error as reported in the table. For orbital inclination, all measured inclinations are reported to be below 90 degrees, as the degeneracy between planets orbiting slightly above and slightly below the plane cannot be broken with photometric measurements. As such, we choose an inclination from within the range of measured errors, then assign this inclination to be above or below 90 degrees with equal probability. This process is repeated independently for each of the inner three planets in each integration. 

The orbital parameters of the outer planet, \thisfourthplanet, are also drawn from the observed values presented in Table \ref{bigtable}. To disentangle $M_{p}$ and $i$, we choose a value for the inclination of \thisfourthplanet\ for each integration, then choose a value of $M_{p} \sin{i}$ from the observed prior, and derive the planetary mass $M_{p}$. 

Each integration results in one of two outcomes: (1) dynamical instabilities, in which planets collide with each other, collide with the central body, or are ejected from the system; or (2) dynamical stability. In this first case of dynamical instability, we assign a probability of those initial conditions reproducing the observations as being 0, as the system loses planets and/or changes orbits significantly. In the second, dynamically stable case, we can perform a second calculation using the results of the numerical simulations and determine the fraction of time that the set of initial conditions in a given integration reproduce the observations. The definition of reproducing the initial conditions requires the three criteria enumerated above.

Each integration results in a pairing of \thisfourthplanet's orbital inclination with a measure of the probability that that particular inclination (and other initial conditions) reproduces the observations. We calculate the average fraction of time {\ron within the trials} in which observations are reproduced as a function of inclination by smoothing the measurements from individual integrations with a Savitzky-Golay filter \citep[a standard low-pass filter;][]{sgfilter}. 

We show the results of our simulations in Figure \ref{inclinationlimits}. In particular, we show the smoothed function and the range of the fraction of time individual realizations of the system satisfy our observational criteria. Evidently, it is hard to reproduce the observed properties of the \thisstar\ system if the outer planet is not aligned close to the inner planets' orbital plane. Unlike the result of \citet{beckeradams}, even a perfectly edge-on system with an inclination of 90$^{\circ}$ does not guarantee that the inner three planets reproduce the observations, mainly because of the increased precision on measured planetary impact parameters.   

Our constraints on the orbital inclination of \thisfourthplanet\ also allow us to place approximate limits on the true mass of this planet {\ron by breaking the degeneracy in the measured $m\sin{i}$ between the planet's mass and orbital inclination}. We find that the 68\% limit on the mass of \thisfourthplanet\ is only 10\% larger than the minimum mass, and that the true mass of \thisfourthplanet\ is smaller than double the minimum mass with 93\% confidence.

\section{Discussion}
\label{discussion}

\begin{figure*}[ht!] 
   \centering
   \includegraphics[scale=0.35]{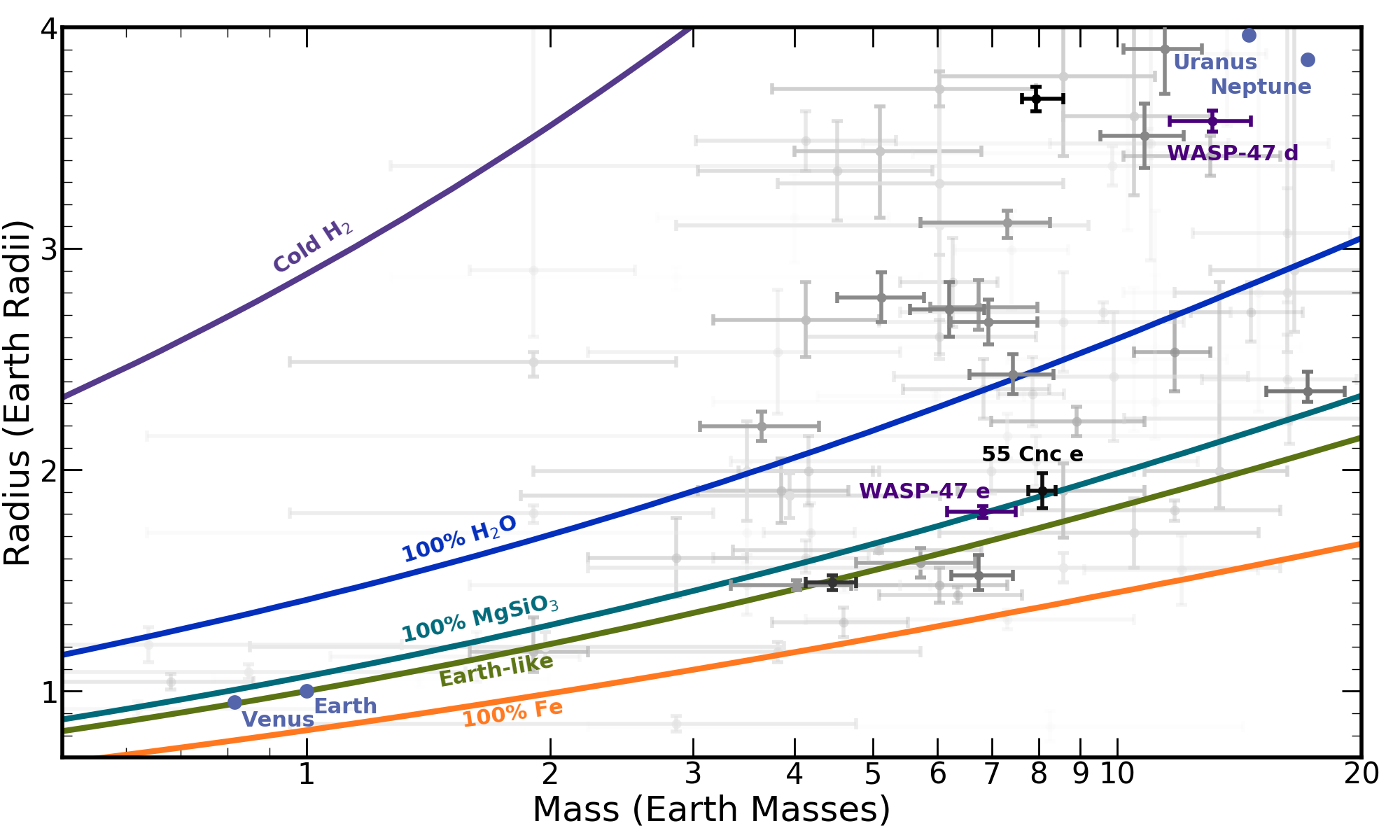} 
   \caption{The mass/radius diagram for small exoplanets. Planet masses and radii are taken from the NASA Exoplanet Archive \citep{akeson}, accessed 2017 Feb 22. The symbol darkness is proportional to the precision with which the masses and radii are determined. Overplotted are theoretical mass/radius relations for solid planets of different compositions from \citet{zeng} and for cold hydrogen planets from \citet{seager}. Solar system planets are shown in blue, and the \thisstar\ planets are shown in purple. We also label 55 Cnc e to show the similarity in composition between it and \thissecondplanet--- both of these planets are less dense than an Earth-like composition, and likely have some volatiles around an Earth-like core.}
   \label{massradius}
\end{figure*}

\subsection{Constraints on the Composition of \thissecondplanet}

Our new measurements and analyses have yielded the most precise planetary masses and radii yet for the \thisstar\ planets, which give us the ability to make strong inferences about the planets' compositions. We downloaded mass-radius relations for planets of various compositions\footnote{The mass/radius relations are available at \url{https://www.cfa.harvard.edu/~lzeng/tables/mrtable3.txt}} from \citet{zeng}, and compared the measured mass (\mple\ $\pm$ \umple\ \mearth) and radius (\rple\ $\pm$ \urple\ \rearth) of \thissecondplanet\ with these relations. We show a mass/radius diagram including the newly determined masses for \thissecondplanet\ and \thisthirdplanet\ in Figure \ref{massradius}. Unlike most other small, highly irradiated planets \citep{dressingk93}, the mass and radius of \thissecondplanet\ are not consistent with an Earth-like composition {\ron (32.5\% iron core, 67.5\% silicate mantle)} at the $p = 5\times 10^{-4}$, or roughly 3.3--$\sigma$ level\footnote{If we instead assume the slightly larger mass from our RV analysis including the PFS data, this conclusion still holds. {\ron Assuming a two-component iron core/rocky silicate mantle model, the median iron fraction for \thissecondplanet\ would be 4.3\% +/- 9.8\%. An Earth-like 32.5\% iron core fraction is excluded at the $p=1.5 \times 10^{-3}$ or 3-$\sigma$ level.}}. Instead, \thissecondplanet\ is less dense than an Earth-like rocky planet, and falls closer to the pure rock (MgSiO$_3$) mass/radius relation from \citet{zeng}.

There are several possibilities for what the composition of \thissecondplanet\ might be. One possibility is that \thissecondplanet\ is a rocky planet with a very small iron core mass fraction compared to the Earth. {\ron Assuming a two-component iron core/rocky silicate mantle model, in this case, we would infer an iron fraction of 1.4\% +/- 8.4\%.} We believe this scenario is unlikely. Theoretically, given the small scatter in chemical abundances of stars in the Solar neighborhood, rocky planet radii should not change much more than 2\% due to differences in compositions \citep{grasset}, while \thissecondplanet's radius is 7\% larger than an Earth-like planet with the same mass. Moreover, observations of small, likely rocky planets near their host stars have shown that rocky exoplanets tend to have compositions consistent with that of the Earth \citep{dressingk93, buchhave16}. It is unclear how a planet of this size, around a star of such high metallicity/iron content, could avoid accumulating any substantial amount of iron. 

Instead, a more likely possibility is that \thissecondplanet\ has an Earth-like core and mantle that is surrounded by a volatile-rich envelope. This type of interior structure is believed to be common among super-Earths and sub-Neptunes discovered by \Kepler\ and K2 since RV observations have shown that most of these planets larger than about 1.6 \rearth\ have densities too low to be explained by rocky compositions {\ron \citep{marcy, weissmarcy, rogers, dressingk93, sinukoffk2106}}. Due to its short (19 hour) orbital period, \thissecondplanet\ is so highly irradiated that any hydrogen/helium envelope would quickly be lost via photo-evaporation \citep{penz2008, sanzforcada, lopez12}, so any envelope around \thissecondplanet\ must be made of water or some other high-metallicity volatile material \citep{lopez}. Using the model described by \citet{lopez}, we find that an Earth-like core and mantle surrounded by a water (or in this case, steam) layer making up 17\% of the \thissecondplanet's total mass is consistent with our observations.


\thissecondplanet\ joins 55 Cnc e as the only ultra-short period (USP, $P<1$ day) planets with densities too low to be consistent with an Earth-like composition\footnote{KOI 1843.03 must have a high density to avoid tidal disruption and therefore a larger iron core mass fraction than Earth \citep{koi1843}.} \citep{sanchisojedausp}. Mass measurements of other transiting USP planets from \Kepler\ and CoRoT have all been consistent with Earth-like compositions  (\Kepler-10 b, \citealt{batalhak10, dumusque, weissk10}; \Kepler-78 b, \citealt{howardk78, pepek78, grunblatt}; CoRoT-7 b, \citealt{queloz, haywood}; K2-106 b, \citealt{sinukoffk2106, guentherk2106}; and HD 3167 b, \citealt{christiansenhd3167, gandolfihd3167}). 

Of all the USP planets with measured masses, \thissecondplanet\ and 55 Cnc e are the two largest and most massive, so perhaps the largest USP planets preferentially retain some volatile materials. However, \thissecondplanet\ and 55 Cnc e are also the two USP planets with the most precise mass determinations --- it is possible that some of the other lower-mass USP planets also have densities too low to be explained by Earth-like compositions, and our data are not yet constraining enough to tell.

How might \thissecondplanet\ and 55 Cnc e have come to possess such compositions? Previously it has been thought that USP planets might commonly be the remains of puffy planets {\ron \citep[but probably not hot Jupiters,][]{winnusp}} after photo-evaporation stripped them of most or all of their volatile envelopes. Evidence for this includes the fact that almost all USP planets or candidates have radii smaller than about twice that of the Earth \citep{jacksonusp, sanchisojedausp, lopez}, in contrast to the large population of less irradiated planets with radii between 2 and 4 \rearth\ at longer orbital periods \citep{fressin, petigura1}. \thissecondplanet\ and 55 Cnc e might therefore be the remnants of larger planets that were massive enough to accrete a significant amount of both hydrogen and denser volatile materials before the hydrogen was subsequently lost to photoevaporation. {\ron On the other hand, WASP-47 e and 55 Cnc e may be unusual -- in addition to being the only known USPs inconsistent with Earth-like interior structures, these planets are also the only well characterized USPs in systems with multiple Jovian planets. If this is not a coincidence and these planets are not typical of USPs, then a more exotic origin scenario may be required. One such possibility is that these planets are the remnant cores of hot Jupiters stripped by Roche lobe overflow \citep{valsecchi, jacksonhjusp}.  Although this mechanism cannot explain the general population of rocky USPs, this could explain \thissecondplanet\ and 55 Cnc e's unusually low densities for such highly irradiated planets, their similar orbital periods (determined by the orbital radius at which Roche lobe overlow began), and the fact that these two objects were found in systems with multiple giant planets.}

Finally, we note several additional similarities between the \thisstar\ system and the 55 Cnc system. Both host stars have high metallicity \citep[\feh\ = \fe\ for \thisstar, and \feh\ = 0.31 for 55 Cnc, ][]{valentifischer}, both systems have ultra-short-period planets which likely have a layer of dense volatile materials (\thissecondplanet\ and 55 Cnc e), both systems have short-period giant planets (\thisfirstplanet\ and the 15d period warm Jupiter 55 Cnc b), and both systems have long-period giant planets (\thisfourthplanet\ and the 5000d period 55 Cnc d). While the 55 Cnc planets may not be as closely aligned with one another as the \thisstar\ planets \citep[only the innermost planet around 55 Cnc is known to transit, and astrometric measurements have shown a misalignment for the giant outer planet 55 Cnc d][]{mcarthur}, the similarities between these two system architectures suggest similar origins. 

\subsection{Constraints on the Composition of \thisthirdplanet}

We also determine a precise mass and radius for \thisthirdplanet\ of \mpld\ $\pm$ \umpld\ \mearth\ and \rpld\ $\pm$ \urpld\ \rearth\, respectively. On a mass/radius diagram (Figure \ref{massradius}), \thisthirdplanet\ is close to, but slightly smaller and less massive than the two Solar system ice giants, Uranus and Neptune. Like Uranus and Neptune, \thisthirdplanet\ must have a low density hydrogen/helium envelope to match our mass and radius measurements, but most of the mass of the planet's mass is in a dense core. Using the models from \citet{lopezsubneptunes}, we find that if the interior has an Earth-like composition with an iron core and rocky mantle, the hydrogen/helium envelope around \thisthirdplanet\ would have a mass fraction of about 5\%. If instead, the solid core is rich in water or other high metallicity volatiles, the hydrogen/helium envelope mass fraction would be closer to 2\%. 

\subsection{Orbital Inclination of \thisfourthplanet}

Even though \thisfourthplanet\ has only been detected in radial velocities, in Section \ref{dynamics}, we were able to put strong constraints on its orbital inclination by requiring dynamical perturbations from its orbit not disrupt the well aligned, co-transiting state of the inner three planets. We found that \thisthirdplanet\ likely has an inclination within a few degrees of an edge-on 90$^\circ$ orbit.

The fact that \thisfourthplanet\ likely orbits in the same plane as the three transiting planets suggests that the system formed in a dynamically quiet manner. If the hot Jupiter, \thisfirstplanet, formed beyond the snow-line and migrated after being scattered by \thisfourthplanet, we might expect the plane of \thisfourthplanet's orbit to be different from the plane of the inner system. Instead, we see a picture more consistent with formation and migration that largely happened in the plane of the protoplanetary disk. In particular, the fact that \thissecondplanet\ likely has a layer of dense volatile material like water suggests that it may have formed beyond the snow-line and migrated to its current location through the protoplanetary disk. \thisfourthplanet's relatively high eccentricity is somewhat difficult to explain in this context, because any eccentricity would have been damped by the disk. In this scenario, the eccentricity must have been excited after the disk dissipated, possibly by another, more distant planet, as suggested by \citet{weissw47}.

\subsection{ \thisfourthplanet's Transit Probability}

Another, more practical, implication of the likely close alignment of \thisfourthplanet\ with the inner transiting system is that the probability of \thisfourthplanet\ transiting is greatly enhanced compared to the naive geometric transit probability. The expression for geometric transit probability is given by \citet{sackett}: 

\begin{equation}
P_{\rm transit, geom} = \frac{\int_{i_{t}}^{90^\circ} \sin{i}\ di}{\int_{0^\circ}^{90^\circ} \sin{i}\ di} 
\label{normal_transit2}
\end{equation}

\noindent where $i$ is the planet's inclination, and $i_t$ is the inclination above which the planet will transit, which depends on the orbital eccentricity $e$, argument of periastron $\omega_p$, semi-major axis $a$, stellar radius $R_\star$, and planetary radius $R_p$ as follows:

\begin{equation}
\cos{i_t} = \frac{R_p + R_\star}{a} \times \frac{1 + e \sin{\omega_p}}{1-e^2}
\label{critical_inclination2}
\end{equation}

We generalized the geometric expression by including a function $\mathcal{P}(i)$, the fraction of time our simulated systems reproduced observations of \thisstar\, inside the integrals, giving: 

\begin{equation}
P_{\rm transit, mod} = \frac{\int_{i_{t}}^{90^\circ} \mathcal{P}(i) \sin{i}\ di}{\int_{0^\circ}^{90^\circ} \mathcal{P}(i) \sin{i}\ di} 
\label{modified_transit2}
\end{equation}

Here, for $\mathcal{P}(i)$ we use the Savitzky-Golay filtered fraction of times that our numerical simulations reproduced observations of \thisstar, and for $i_{t}$ we use 89.66$^\circ$. 

Without {\em a priori} knowledge of our inclination constraints for \thisfourthplanet, the transit probability is about 0.6\%. When we take into account our dynamical inclination constraints, the probability increases by more than an order of magnitude to about 10\% -- the same {\em a priori} transit probability of a typical hot Jupiter. 

If \thisfourthplanet\ is found to transit, it would open the door to future sophisticated investigations into the properties and formation history of the \thisstar\ planets. It would be possible to study the atmosphere of both \thisfirstplanet\ and \thisfourthplanet\ in transit with the upcoming {\em James Webb Space Telescope}, determine and compare atmospheric abundances, and infer these planets' birthplaces \citep{oberg}. A detection of a transit of \thisfourthplanet\ could make the \thisstar\ system a key to unlocking the origin of hot Jupiters. 

Detecting a transit of \thisfourthplanet\ should not be difficult - the transit depth would likely be about 1\%, easily attainable by ground-based telescopes with moderate apertures. The transit duration will be long - an equatorial transit of \thisfourthplanet\ would last 14 hours, so a successful detection would likely require a coordinated ground-based campaign with several telescopes longitudally dispersed around the globe. At present, the largest obstacle to successfully recovering a transit of \thisfourthplanet\ is the uncertainty in the transit time. The last transit window happened around 9 January 2017, while \thisstar\ was unobservable behind the Sun, with an uncertainty in transit time of 4.8 days. The next several transit windows will be around 21 August 2018 $\pm$ 6.5 days, 31 March 2020 $\pm$ 8.6 days, and 10 November 2021 $\pm$ 10.8 days. It would take a massive ground-based campaign to cover enough of these transit windows to ensure success. Although the uncertainties on the transit times are large now, they will sharpen considerably once precise RV spectographs have completed observing a full orbital period of \thisfourthplanet. We will continue to observe \thisstar\ with HARPS-N in the coming years to refine the orbital period and ephemeris of \thisfourthplanet\ in preparation for the chance to detect the planet in transit. 

\section{Summary}
\label{summary} 

We have investigated the \thisstar\ planetary system, which is known to host a hot Jupiter, two smaller transiting planets flanking the hot Jupiter, and a long-period Jovian companion. Using new data from the HARPS-N spectrograph and previously published data from the K2 mission and other ground-based spectrographs, we have measured the masses and radii of the transiting planets, and determined the orbit of the outer planet. Our main conclusions are summarized as follows: 

\begin{enumerate}
\item We have measured the most precise masses and radii for the \thisstar\ planets yet. The innermost planet, \thissecondplanet, has a mass of  \mple\ $\pm$ \umple\ \mearth\ and a radius \rple\ $\pm$ \urple\ \rearth. The hot Jupiter, \thisfirstplanet\ has mass \mplb\ $\pm$ \umplb\ \mearth\ and a radius \rplb\ $\pm$ \urplb\ \rearth. We find the Neptune-sized planet, \thisthirdplanet, has a mass \mpld\ $\pm$ \umpld\ \mearth\ and radius \rpld\ $\pm$ \urpld\ \rearth. The outer Jovian planet, \thisfourthplanet\ is not known to transit, so from our radial velocity observations, we only measure the planet's minimum mass $m\sin{i}$ of \mplc\ $\pm$ \umplc\ \mearth. 
\item \thissecondplanet, unlike most other planets in ultra short period orbits, does not have an Earth-like composition. We find that \thissecondplanet\ is not dense enough to have an iron core with the same mass fraction as terrestrial planets in the Solar system. Instead, \thissecondplanet\ likely has a volatile rich (possibly water/steam) envelope comprising 17\% its total mass on top of an Earth-like core. 
\item We show using dynamical simulations that the inclination of \thisfourthplanet\ is likely well aligned with the inner transiting system. The orbital inclination of \thisfourthplanet\ is likely within a few degrees of edge on in order to not disrupt the inner transiting planets from their present-day well aligned configuration. This alignment, plus the alignment between the planets' orbits and the stellar spin axis \citep{sanchisojedaw47rm} suggests a dynamically quiet formation/migration scenario for the \thisstar\ planets that kept all of the planets in the plane of the protoplanetary disk. The outer planet is much more likely to transit than the geometric transit probability, motivating campaigns to observe the transit in future opportunities. Additionally, this limit on the inclination suggests that the true mass of the \thisfourthplanet\ is likely close to the measured $M_P\sin{i}$.
\end{enumerate}

Future radial velocity observations of \thisstar\ will both continue to improve the precision on the masses of the two smaller planets, and will greatly improve the precision on the predicted transit time of \thisfourthplanet. Sharpening the transit predictions will be hugely important to making a campaign to detect or rule out transits of \thisfourthplanet\ feasible.

\acknowledgments
We thank Jason Eastman, Jonathan Irwin, Laura Kreidberg, Willie Torres, Lauren Weiss, and George Zhou for helpful conversations. We thank Josh Winn for helpful comments on the manuscript and the anonymous referee for a thoughtful report. A.V. and J.C.B are supported by the NSF Graduate Research Fellowship, grant nos. DGE 1144152 and DGE 1256260, respectively. This work was performed in part under contract with the California Institute of Technology/Jet Propulsion Laboratory funded by NASA through the Sagan Fellowship Program executed by the NASA Exoplanet Science Institute. D.W.L. acknowledges partial support from the from the TESS mission through a sub-award from the Massachusetts Institute of Technology to the Smithsonian Astrophysical Observatory. The research leading to these results has received funding from the European Union Seventh Framework Programme (FP7/2007-2013) under Grant Agreement n. 313014 (ETAEARTH). Parts of this work have been supported by NASA under grants No. NNX15AC90G and NNX17AB59G issued through the Exoplanets Research Program. This publication was made possible through the support of a grant from the John Templeton Foundation. The opinions expressed in this publication are those of the authors and do not necessarily reflect the views of the John Templeton Foundation.

This work is based on observations made with the Italian Telescopio Nazionale Galileo (TNG) operated on the island of La Palma by the Fundación Galileo Galilei of the INAF (Istituto Nazionale di Astrofisica) at the Spanish Observatorio del Roque de los Muchachos of the Instituto de Astrofisica de Canarias. The HARPS-N project was funded by the Prodex Program of the Swiss Space Office (SSO), the Harvard University Origin of Life Initiative (HUOLI), the Scottish Universities Physics Alliance (SUPA), the University of Geneva, the Smithsonian Astrophysical Observatory (SAO), and the Italian National Astrophysical Institute (INAF), University of St. Andrews, Queens University Belfast and University of Edinburgh. This work was supported in part by the NASA Exoplanets Research Program. 
 
This work used the Extreme Science and Engineering Discovery Environment (XSEDE), which is supported by National Science Foundation grant number ACI-1053575. This research was done using resources provided by the Open Science Grid, which is supported by the National Science Foundation and the U.S. Department of Energy's Office of Science. We have made use of NASA's Astrophysics Data System and the NASA Exoplanet Archive, which is operated by the California Institute of Technology, under contract with the National Aeronautics and Space Administration under the Exoplanet Exploration Program. 

This paper includes data collected by the \Kepler/K2 mission. Funding for the \Kepler\ mission is provided by the NASA Science Mission directorate. Some of the data presented in this paper were obtained from the Mikulski Archive for Space Telescopes (MAST). STScI is operated by the Association of Universities for Research in Astronomy, Inc., under NASA contract NAS5--26555. Support for MAST for non--HST data is provided by the NASA Office of Space Science via grant NNX13AC07G and by other grants and contracts.

Facilities: \facility{Kepler/K2, TNG (HARPS-N)}


\begin{deluxetable*}{lcccc}
\tablecaption{System Parameters for \thisstar \label{bigtable}}
\tablewidth{0pt}
\tablehead{
  \colhead{Parameter} & 
  \colhead{Value}     &
  \colhead{} &
  \colhead{68.3\% Confidence}     &
  \colhead{Comment}   \\
  \colhead{} & 
  \colhead{}     &
  \colhead{} &
  \colhead{Interval Width}     &
  \colhead{}  
}
\startdata
\emph{Stellar Parameters} & & & \\
Right Ascension & 22:04:48.7 & & &  \\
Declination & -12:01:08 & & &  \\

$M_\star$~[$M_\odot$] & \mst & $\pm$&$ \umst$ & A,B \\
$R_\star$~[$R_\odot$] & \rst & $\pm$&$ \urst$ & A,B \\
Limb darkening $q_1$~ & \ldone  & $\pm$ &$ \uldone$ & B \\
Limb darkening $q_2$~ & \ldtwo  & $\pm$ &$ \uldtwo$ & B \\
Stellar Density$\rho_\star$~[$g\,cm^{-3}$] & \rhost & $\pm$&$ \urhost$ & B \\

$\log g_\star$~[cgs] & \loggst & $\pm$&$ \uloggst$ & A,B,C \\
\meh & $\fe$ & $\pm$&$ \ufe$ & C \\
$T_{\rm eff}$ [K] & \teff & $\pm$&$ \uteff$ & C\\
 & & \\
 
\emph{\thissecondplanet} & & & \\
Orbital Period, $P$~[days] & \perple & $\pm$&$ \uperple $ & B \\
Radius Ratio, $(R_P/R_\star)$ & \rprste & $\pm$&$ \urprste$ & B \\
Scaled semi-major axis, $a/R_\star$  & \arste & $\pm$&$ \uarste$ & B \\
Orbital inclination, $i$~[deg] & \incle & $\pm$&$ \uincle$ & B \\
Transit impact parameter, $b$ & \impe & $\pm$&$ \uimpe$ & B \\
Time of Transit $t_{t}$~[$\rm BJD_{\rm TDB}$] & \ttransite & $\pm$& \uttransite & B\\ 
Transit Duration $t_{14}$~[hours] & \tdure & $\pm$& \utdure & B\\ 
RV Semiamplitude $K_e$~[\ms] & \ke & $\pm$& \uke & \ron{D}\\ 
$M_P$~[\mearth] & $\mple$  &  $\pm$ & $\umple$  & A,D \\
$R_P$~[\rearth] & $\rple$ &   $\pm$&$ \urple$  & A,B \\
Surface Gravity~[\mssq] & \gpe &   $\pm$&$ \ugpe$  & A,B,D \\
Mean Density~[\gcc] & \rhoe &   $\pm$&$ \urhoe$  & A,B,D \\

 & & \\

\emph{\thisfirstplanet} & & & \\
Orbital Period, $P$~[days] & \perplb & $\pm$&$ \uperplb $ & B \\
Radius Ratio, $(R_P/R_\star)$ & \rprstb & $\pm$&$ \urprstb$ & B \\
Scaled semi-major axis, $a/R_\star$  & \arstb & $\pm$&$ \uarstb$ & B \\
Orbital inclination, $i$~[deg] & \inclb & $\pm$&$ \uinclb$ & B \\
Transit impact parameter, $b$ & \impb & $\pm$&$ \uimpb$ & B \\
Time of Transit $t_{t}$~[$\rm BJD_{\rm TDB}$] & \ttransitb & $\pm$& \uttransitb & B\\
Transit Duration $t_{14}$~[hours] & \tdurb & $\pm$& \utdurb & B\\ 
RV Semiamplitude $K_b$~[\ms] & \kb & $\pm$& \ukb & D\\ 
$M_P$~[\mearth] & \mplb  &   & \umplb & A,D \\
$R_P$~[\rearth] & \rplb &   $\pm$&$ \urplb$  & A,B \\
Surface Gravity~[\mssq] & \gb &   $\pm$&$ \ugb$  & A,B,D \\
Mean Density~[\gcc] & \rhob &   $\pm$&$ \urhob$  & A,B,D \\
 & & \\

\emph{\thisthirdplanet} & & & \\
Orbital Period, $P$~[days] & \perpld & $\pm$&$ \uperpld $ & B \\
Radius Ratio, $(R_P/R_\star)$ & \rprstd & $\pm$&$ \urprstd$ & B \\
Scaled semi-major axis, $a/R_\star$  & \arstd & $\pm$&$ \uarstd$ & B \\
Orbital inclination, $i$~[deg] & \incld & $\pm$&$ \uincld$ & B \\
Transit impact parameter, $b$ & \impd & $\pm$&$ \uimpd$ & B \\
Time of Transit $t_{t}$~[$\rm BJD_{\rm TDB}$] & \ttransitd & $\pm$& \uttransitd & B\\ 
Transit Duration $t_{14}$~[hours] & \tdurd & $\pm$& \utdurd & B\\ 

RV Semiamplitude $K_d$~[\ms] & \kd & $\pm$& \ukd & D\\ 
$M_P$~[\mearth] & \mpld &   & \umpld & A,D \\
$R_P$~[\rearth] & \rpld &   $\pm$&$ \urpld$  & A,B \\
Surface Gravity~[\mssq] & \gd &   $\pm$&$ \ugd$  & A,B,D \\
Mean Density~[\gcc] & \rhod &   $\pm$&$ \urhod$  & A,B,D \\
Eccentricity & $< 0.014$ &    &  & E \\
 & & \\

\emph{\thisfourthplanet} & & & \\
Orbital Period, $P$~[days] & \perplc & $\pm$&$ \uperplc $ & D \\
Time of Inferior Conjunction $t_{t}$~[$\rm BJD_{\rm TDB}$] & \ttransitc & $\pm$& \uttransitc & D\\ 
$M_P\sin{i}$~[\mearth] & \mplc &   & \umplc & A,D \\
Eccentricity & \eccc &   $\pm$&$ \ueccc$  & D \\
Argument of Periastron~[degrees] & \omegac &   $\pm$&$ \uomegac$  & D \\
Semimajor Axis~[AU] & \ac &   $\pm$&$ \uac $ & A,D \\

 & & \\

\enddata

\tablecomments{A: Parameters come from our stellar parameter analysis in Section \ref{sec:stellar_prop}. B: Parameters come from analysis of the K2 light curve in Section \ref{sec:lc_analysis}. C. Parameters come from weighted average of spectroscopic parameters from three different methods described in Section \ref{sec:spec_prop}. D: Parameters come from our radial velocity analysis in Section \ref{sec:rv_analysis}. E: The eccentricity of \thisthirdplanet\ was fit with a strong Gaussian prior of 0 $\pm$ 0.014 from TTV and dynamical stability arguments. The argument of periastron was not constrained in our fits either by the data or prior.}

\end{deluxetable*}

\clearpage

\end{document}